\documentclass[12pt]{revtex4}%
\usepackage{amsfonts}
\usepackage{amsmath}
\usepackage{amssymb}
\usepackage{graphicx}
\usepackage{color}
\usepackage{epsfig}%
\providecommand{\U}[1]{\protect\rule{.1in}{.1in}}

\begin{document}
\title{Stochastic Variational Method as a Quantization Scheme II: Quantization of
Electromagnetic Fields}
\author{T. Koide}
\author{T. Kodama}
\affiliation{Instituto de F\'{\i}sica, Universidade Federal do Rio de Janeiro, C.P. 68528,
21941-972, Rio de Janeiro, Brazil}
\author{K. Tsushima}
\affiliation{International Institute of Physics, Federal University of Rio Grande do Norte,
Natal 59078-400, RN, Brazil}

\begin{abstract}
Quantization of electromagnetic fields is investigated in the framework of
stochastic variational method (SVM). Differently from the canonical
quantization, this method does not require canonical form and quantization can
be performed directly from the gauge invariant Lagrangian. The gauge condition
is used to choose dynamically independent variables. We verify that, in the
Coulomb gauge condition, SVM result is completely equivalent to the
traditional result. On the other hand, in the Lorentz gauge condition, SVM
quantization can be performed without introducing the indefinite metric. The
temporal and longitudinal components of the gauge filed, then, behave as
c-number functionals affected by quantum fluctuation through the interaction
with charged matter fields. To see further the relation between SVM and the
canonical quantization, we quantize the usual gauge Lagrangian with the Fermi
term and argue a stochastic process with a negative second order correlation
is introduced to reproduce the indefinite metric.

\end{abstract}

\maketitle

\section{Introduction}

\begin{figure}[t]
\includegraphics[scale=0.4]{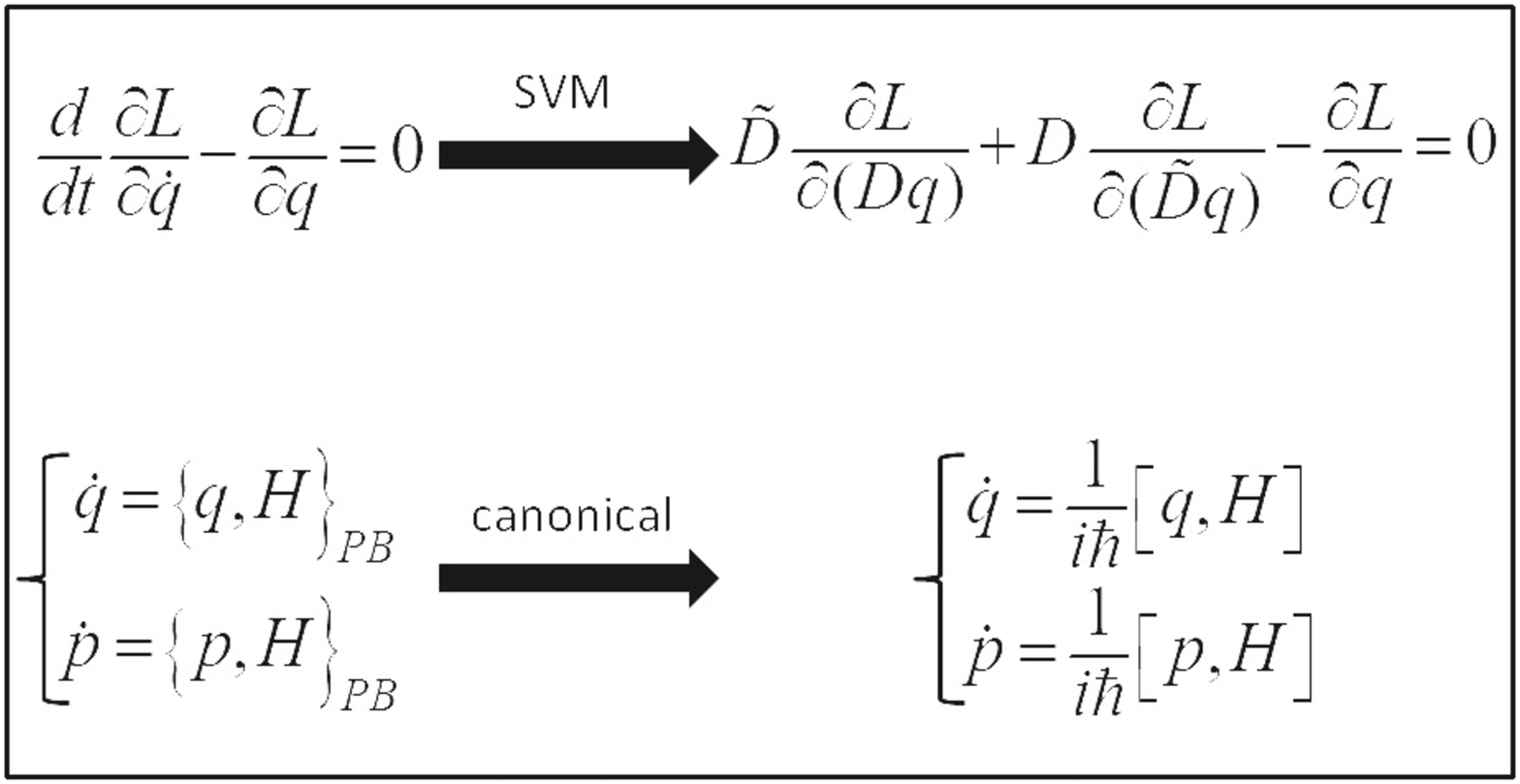}\caption{For quantization, the
Euler-Lagrange equation is modified in SVM, while the canonical equation is
changed in the canonical quantization. The mean forward and backward
derivatives respectively $D$ and $\tilde{D}$ are defined in Ref. \cite{kk4}.}%
\label{fig1}%
\end{figure}

Stochastic variational method (SVM) is a generalized method of the variational
procedure where dynamical variables are extended to the domain of stochastic
ones. Instead of determining an optimal path of each stochastic process, this
variational method aims to determine the optimized evolution of the
probability distribution function. The method was firstly introduced by Yasue
\cite{yasue} to reformulate Nelson's stochastic quantization method
\cite{nelson1,nelson2} from the point of view of the variation principle.

Subsequently, SVM has been extended to more general cases. For example, the
Navier-Stokes equation \cite{kk1,kk2,koide}, the Gross-Pitaevskii equation
\cite{morato,kk2}, and the Schr\"{o}dinger-Langevin (Kostin) equation \cite{misawa} are formulated in SVM
approach. Further aspects of SVM have also been studied, such as the Noether's
theorem \cite{misawa3}, the uncertainty relation \cite{kk3}, the applications
to the many-body particle systems \cite{morato}, and continuum media
\cite{kk1,kk2}. Although many groups
\cite{dav,guerra,zamb,hase,marra,jae,pav,rosen,wang,naga,cre,arn,kappen,eyink,gomes,hiroshima,serva,yamanaka}
have investigated SVM, the applicability has not yet been well explored. Thus
it is worth investigating the applicability of SVM to more complex systems.

In this series of works, we focus on the applicability of SVM as an
alternative field quantization scheme. In Part I, we discussed the formulation
of SVM to the field quantization and showed that the complex Klein-Gordon
equation can be quantized appropriately and this method has an advantage for
the definition of the Noether charge \cite{kk4}. This paper is Part II and we
discuss the quantization of electromagnetic fields in the framework of SVM.

There is a sufficient reason to expect that the SVM quantization provides
another perspective in the gauge field quantization compared to the usual
canonical quantization. The canonical quantization is implemented by employing
the commutation relations of canonically conjugate variables. As is
well-known, however, this procedure is not straightforward for electromagnetic
fields because the canonical momentum for the time component of the
four-vector gauge field $A^{\mu} (x)$ vanishes. To circumvent this difficulty
in a covariant manner, for example, the so-called Fermi term is introduced to
the Lagrangian density, paying the price of introducing the indefinite metric
(Fock state vector with a negative norm). We need to take a special care to
project out the undesirable negative norm states from the physical Fock space.

Such a difficulty of the canonical formulation with dynamical constraints
already appears in the classical level. Nevertheless, this does not give rise
to any problem to derive the classical equations of motion by the variational
procedure. Therefore, if quantization can be regarded as the stochastic
generalization of the optimization of actions as is claimed in SVM, the gauge
field quantization should be performed directly to the same classical action
without introducing additional terms. See also Fig. \ref{fig1}.

The principal purposes of the present work are twofold. One is to confirm our
speculation mentioned above, that is, to apply the SVM quantization to the
gauge invariant Lagrangian and discuss the properties of the derived quantized
dynamics. The other is to investigate the counterpart of the indefinite metric
in SVM. If SVM gives the consistent framework of quantization, it will be
applicable even to the usual gauge Lagrangian with the Fermi term and the
well-known result should be reproduced. For this, we need to extend the
concept of a stochastic process to represent the indefinite metric in SVM.

This paper is organized as follows. In Sec. \ref{sec:lag}, for the sake of
book-keeping, we introduce our notation for the discretization scheme of the
Lagrangian density of electromagnetic fields. In Sec. \ref{sec:invariant}, the
stochastic variation is applied to the gauge invariant Lagrangian density. In
Sec. \ref{sec:fix} the application to the Lagrangian density with the Fermi term 
is discussed by generalizing the concept of stochastic process to
reproduce the indefinite metric. Section \ref{sec:conc} is devoted to
concluding remarks.

\section{Stochastic Lagrangian Density of Electromagnetic Fields}

\label{sec:lag}

In SVM, quantum fluctuation is introduced as random noise in a classical
system. Thus, for a system of fields, the field configuration becomes not
smooth in space and time. To deal with such a behavior, we introduce the space
lattice discretization of the field variables as is done in \cite{kk4} . In
this section, for the sake of book-keeping of notations, we summarize the
scheme, extending to the vector fields.

\subsection{Discretized Expression of Derivatives}

For the discretization scheme, we introduce a set of $N^{3}$ cubic lattice
grid points forming a cubic domain of side $L$, volume $V=L^{3}.$ The side of
the unit lattice cube $\Delta x$ is $\Delta x$ $=L/N$. We denote the set of
the whole lattice point by $\Omega=\left\{  \mathbf{x}=(x_{l},y_{m}%
,z_{n})=(l\Delta x,m\Delta x,n\Delta x),\ 0\leq l,m,n\leq N-1\right\}  $. For
a given time, spatial configuration of a field is completely specified by a
set of values $f=\{f_{\mathbf{x}};\mathbf{x\in}\Omega\}$ imposing the periodic
boundary condition in each direction. For this purpose, $N$ is ought to be an
even integer. The periodic boundary condition is then expressed as
\begin{equation}
f_{\mathbf{x}}=f_{\mathbf{x}+\mathbf{e}^{i}L},
\end{equation}
where $\mathbf{e}^{i}$ is defined by $\mathbf{e}^{x}=(1,0,0)$, $\mathbf{e}%
^{y}=(0,1,0)$ and $\mathbf{e}^{z}=(0,0,1)$. Then the spatial derivative is
defined by
\begin{equation}
\nabla^{i}f_{\mathbf{x}}=\frac{f_{\mathbf{x}+\mathbf{e}^{i}\Delta
x}-f_{\mathbf{x}-\mathbf{e}^{i}\Delta x}}{2\Delta x},
\end{equation}
and the corresponding Laplacian operator is expressed as
\begin{equation}
\Delta_{\mathbf{x}}f_{\mathbf{x}}\equiv\sum_{i}\nabla^{i}\nabla^{i}%
f_{\mathbf{x}}=\sum_{i}\frac{f_{\mathbf{x}+2\mathbf{e}^{i}\Delta
x}-2f_{\mathbf{x}}+f_{\mathbf{x}-2\mathbf{e}^{i}\Delta x}}{4(\Delta x)^{2}}.
\end{equation}
Note that the spatial derivative here is the average of the two spatial 
derivatives defined Ref. \cite{kk4}. This is to simplify the introduction of
the gauge conditions. We verify the partial integration formula over the whole
space as in Refs. \cite{kk1,kk2,kk4},
\begin{equation}
\sum_{\mathbf{x\in\Omega}}h_{\mathbf{x}}\nabla g_{\mathbf{x}}=-\sum
_{\mathbf{x\in\Omega}}g_{\mathbf{x}}\nabla h_{\mathbf{x}}%
\end{equation}
for any field configurations $f$ and $g$, due to the periodic boundary condition.

On the other hand, we keep the two different discretized definitions for the
time derivative as,
\begin{subequations}
\begin{align}
(\partial_{t})_{+}f_{\mathbf{x}}(t)  &  =\frac{f_{\mathbf{x}}%
(t+dt)-f_{\mathbf{x}}(t)}{dt}~~~~(dt\rightarrow0_{+}),\label{d+}\\
(\partial_{t})_{-}f_{\mathbf{x}}(t)  &  =\frac{f_{\mathbf{x}}%
(t+dt)-f_{\mathbf{x}}(t)}{dt}~~~~(dt\rightarrow0_{-}). \label{d-}%
\end{align}
\end{subequations}
These do not coincide even in the limit $\left(  \left\vert dt\right\vert
\rightarrow0\right) $ in the presence of noises, as is discussed in
Refs.~\cite{kk2,kk4}. Note that SVM is formulated based on the stochastic
calculus of the Wiener process and thus any higher order contribution in terms
of $dt$ should be ignored. For the space direction, the infinitesimal limit of
$\Delta x$ is taken in the end of calculations.

For the later convenience, we introduce the following notations,
\begin{equation}
(\partial_{+})_{\mu}=\left(  \frac{1}{c}(\partial_{t})_{+},\nabla\right)
,~~~~(\partial_{-})_{\mu}=\left(  \frac{1}{c}(\partial_{t})_{-},\nabla\right)
,
\end{equation}
where $(\partial_{\pm})^{\mu}\equiv g^{\mu\nu}(\partial_{\pm})_{\nu}$, $c$ is
the speed of light and $g^{\mu\nu}=diag(1,-1,-1,-1)$ is the Minkowsky metric
tensor. These are just convenient notations and do not mean that these are
Lorentz covariant.

\subsection{Discretized Expression of Gauge Transform}

Let us now introduce the gauge field in the discretized form as
\begin{align}
A^{\mu} (x)=(A^{0}(\mathbf{x},t), \mathbf{A}(\mathbf{x},t) ) \longrightarrow
A^{\mu}_{\mathbf{x}} (t) = (A^{0}_{\mathbf{x}}(t), \mathbf{A}_{\mathbf{x}}(t)
).
\end{align}

From the definitions introduced in the previous subsection, the gauge
transform can be expressed in two different ways;
\begin{subequations}
\begin{align}
A_{\mathbf{x}}^{\mu}(t)\longrightarrow(A_{\mathbf{x}}^{\prime})^{\mu}(t)  &
=A_{\mathbf{x}}^{\mu}(t)-\partial_{+}^{\mu}\lambda_{\mathbf{x}}(t),\label{gt1}%
\\
A_{\mathbf{x}}^{\mu}(t)\longrightarrow(A_{\mathbf{x}}^{\prime})^{\mu}(t)  &
=A_{\mathbf{x}}^{\mu}(t)-\partial_{-}^{\mu}\lambda_{\mathbf{x}}(t).
\label{gt2}%
\end{align}
\end{subequations}
Here $\lambda_{\mathbf{x}}(t)$ is an arbitrary smooth function in time.
Therefore the above two expressions of the gauge transforms are completely
equivalent since $\partial_{+}^{\mu}\lambda_{\mathbf{x}}(t)=\partial_{-}^{\mu
}\lambda_{\mathbf{x}}(t)$ in the infinitesimal $\left\vert dt\right\vert $
limit as mentioned above.

Then, by applying the argument in Ref. \cite{kk4}, the discretized Lagrangian
density is given by the average of the $+$ and $-$ contributions as
\begin{equation}
\mathcal{L}=-\frac{1}{8}F_{+}^{\mu\nu}(F_{+})_{\mu\nu}-\frac{1}{8}F_{-}%
^{\mu\nu}(F_{-})_{\mu\nu}, \label{cla-lag}%
\end{equation}
where
\begin{subequations}
\begin{align}
F_{+}^{\mu\nu}  &  = \partial_{+}^{\mu}A^{\nu}-\partial_{+}^{\nu}A^{\mu
},\label{Fmunu+}\\
F_{-}^{\mu\nu}  &  = \partial_{-}^{\mu}A^{\nu}-\partial_{-}^{\nu}A^{\mu}.
\label{Fmunu-}%
\end{align}
\end{subequations}
See also Refs. \cite{kk2,kk3}. It should be stressed that this Lagrangian
density is invariant under the gauge transforms Eq.~(\ref{gt1}) and Eq.
(\ref{gt2}) even for a finite spatial grid.

As is well-known, discretization scheme of the gauge field has been already
formulated in terms of the link variables and widely used in the lattice field
theory. In the present work, however, we will not introduce the link variable
because the SVM quantization results are compared directly to that of
canonical quantization of electromagnetic fields.

\subsection{$\mathbf{k}-$Representation and Polarization Vector}

To employ the stochastic variation, we have to specify independent degrees of
freedom, to each of which independent noises are introduced. In the present
case, the Lagrangian density~(\ref{cla-lag}) is expressed in terms of the
gauge field $A^{\mu}$ which has four components, but only two of them are
independent due to the gauge invariance.

For this purpose, it is natural to choose the two transverse components as the
two independent variables, because the classical electromagnetic wave contains
only the transverse components. The transverse component in the discretized
form is defined by
\begin{equation}
\nabla\cdot\mathbf{A}_{\bot}=0, \label{Transverse}%
\end{equation}
where the subscript $\bot$ denotes the transverse component.

As was discussed in Ref.~\cite{kk4}, the field quantization in SVM can be done
both in the $\mathbf{x}-$ and $\mathbf{k}-$representations. The properties of
the quantized electromagnetic fields is studied extensively in the momentum
space. Thus, in the following, we develop SVM quantization in the
$\mathbf{k}-$representation.

The field variables in the $\mathbf{k}-$representation, $A_{\mathbf{k}}^{(r)}%
$, is defined by the following linear transform,
\begin{equation}
\mathbf{A}_{\mathbf{x}}=\sqrt{\Delta^{3}\mathbf{k}}\sum_{r=1}^{3}%
\sum_{\mathbf{k}}\mathbf{e}_{\mathbf{k},r}A_{\mathbf{k}}^{(r)}(t)\frac
{e^{i\mathbf{k\cdot x}}}{\sqrt{V}},
\end{equation}
where $\Delta^{3}\mathbf{k}=(2\pi)^{3}/L^{3}$ and the three vectors $\left\{
\mathbf{e}_{\mathbf{k},r},r=1,2,3\right\}  $ are orthogonal unit vectors,
forming an orthonormal base for each given $\mathbf{k}$. Now, we can always
choose that
\begin{equation}
\mathbf{e}_{\mathbf{k},3}=\mathbf{q}_{\mathbf{k}}/\left\vert \mathbf{q}%
_{\mathbf{k}}\right\vert ,\ \label{e3=q}%
\end{equation}
where $\mathbf{q}_{\mathbf{k}}$ is the real eigenvalue of the discretized
derivative operator $\nabla$ with periodic boundary conditions, satisfying
\begin{equation}
\nabla\ e^{i\mathbf{k\cdot x}}=i\mathbf{q}_{\mathbf{k}}\ e^{i\mathbf{k\cdot
x}}, \label{Eigen_D}%
\end{equation}
and expressed as
\begin{equation}
\mathbf{q}_{\mathbf{k}}=\frac{1}{\Delta x}(\sin(k_{x}\Delta x),\sin
(k_{y}\Delta x),\sin(k_{z}\Delta x)), \label{qk}%
\end{equation}
where $\mathbf{k}=\frac{2\pi}{L}(n_{x},n_{y},n_{z})$ with $n_{i}~(i=x,y,z)$
being an integer satisfying $-(N-1)/2\leq n_{i}\leq(N-1)/2$. Note that, as was
discussed in Ref.~\cite{kk4}, $\mathbf{q}_{\mathbf{k}}$ is reduced to
$\mathbf{k}$ in the continuum limit.

From the completeness of 
$\left\{  \mathbf{e}_{\mathbf{k},r},r=1,2,3\right\}$, it is obvious that
\begin{equation}
\sum_{r=1,2}\mathbf{e}_{\mathbf{k},r}\mathbf{e}_{\mathbf{k},r}^{T}%
=I-\frac{\mathbf{q}_{\mathbf{k}}\ \mathbf{q}_{\mathbf{k}}^{T}}{\mathbf{q}%
_{\mathbf{k}}^{2}},
\end{equation}
where $I$ is $\left(  3\times3\right)  $ an identity matrix and the
superscript $^{T}$ represents the transpose operation. Then the two vectors
$\left\{  \mathbf{e}_{\mathbf{k},r},r=1,2\right\} $ are identified as the
polarization vectors. To implement the real property of the field, we use the
convention normally adopted,%
\begin{subequations}
\begin{align}
\mathbf{e}_{-\mathbf{k},r}  &  = \mathbf{e}_{\mathbf{k},r}~~~~(r=1,2
),\label{e12(-k)}\\
\mathbf{e}_{-\mathbf{k},3}  &  = -\mathbf{e}_{\mathbf{k},3} .\label{e3(-k)}%
\end{align}
\end{subequations}

In terms of these polarization vectors, the transverse field 
$\mathbf{A}_{\bot}$ can simply expressed as%
\begin{equation}
\mathbf{A}_{\bot\mathbf{x}}=\sqrt{\Delta^{3}\mathbf{k}}\sum_{r=1}^{2}%
\sum_{\mathbf{k}}\mathbf{e}_{\mathbf{k},r}A_{\mathbf{k}}^{(r)}(t)\frac
{e^{i\mathbf{k\cdot x}}}{\sqrt{V}}. \label{Atrans=k}%
\end{equation}
On the other hand, the temporal and longitudinal components $A_{\mathbf{x}%
}^{0}(t)$ and $\mathbf{A}_{\Vert\mathbf{x}}(t)$, which are not the dynamical
variables in the present case, are expressed as
\begin{align}
A_{\mathbf{x}}^{(0)}  &  =\sqrt{\Delta^{3}\mathbf{k}}\sum_{\mathbf{k}%
}A_{\mathbf{k}}^{(0)}(t)\frac{e^{i\mathbf{k\cdot x}}}{\sqrt{V}},\\
\mathbf{A}_{\Vert\mathbf{x}}  &  =\sqrt{\Delta^{3}\mathbf{k}}\sum_{\mathbf{k}%
}\mathbf{e}_{\mathbf{k},3}A_{\mathbf{k}}^{(3)}(t)\frac{e^{i\mathbf{k\cdot x}}%
}{\sqrt{V}}.
\end{align}

Note that our new independent variables 
$\left\{  A_{\mathbf{k}}^{(r)}(t),A_{-\mathbf{k}}^{(r)}(t),r=1,2\right\}  $ 
are complex but not
independent, because $A_{\mathbf{k}}^{(r)\ast}(t)=A_{-\mathbf{k}}^{(r)}(t)$
because the condition that $\mathbf{A}_{\bot\mathbf{x}}$ is a real field with
the convention Eq.(\ref{e12(-k)}). In terms of real and imaginary parts,
\begin{equation}
A_{\mathbf{k}}^{(r)}(t) =\frac{R_{\mathbf{k}}^{(r)}(t)+i I_{\mathbf{k}}%
^{(r)}(t)}{\sqrt{2}}, \label{eqn:rea-ima}%
\end{equation}
with
\begin{equation}
R_{\mathbf{k}}^{(r)}(t)= R_{\mathbf{-k}}^{(r)}(t), ~~~~ I_{\mathbf{k}}%
^{(r)}(t)= - I_{\mathbf{-k}}^{(r)}(t).
\end{equation}

For the sake of the later convenience, we introduce a new real variables,
$\left\{  \xi_{\mathbf{k}}^{(r)},r=1,2\right\}  $ for any $\mathbf{k}$ from
which we define the field amplitudes as
\begin{subequations}
\label{xi}
\begin{align}
\xi_{\mathbf{k}}^{(r)}(t)  &  = R_{\mathbf{k}}^{(r)}(t) ,\label{ReA}\\
\xi_{\mathbf{-k}}^{(r)}(t)  &  = I_{\mathbf{k}}^{(r)}(t). \label{ImA}%
\end{align}
\end{subequations}

\section{Stochastic Variation for Gauge Invariant Lagrangian}

\label{sec:invariant}

\subsection{transverse components}

We apply the stochastic variational method to the gauge invariant Lagrangian
density (\ref{cla-lag}) following Ref. \cite{kk4}.

Now we replace these fields with stochastic variables as
\begin{equation}
\xi_{\mathbf{k}}^{(r)}(t)\rightarrow\hat{\xi}_{\mathbf{k}}^{(r)}(t).
\end{equation}
Here the symbol $\symbol{94}$ is used to express stochastic variables. The
forward and backward stochastic differential equations (SDEs) are,
respectively, given by
\begin{subequations}
\begin{align}
d\hat{\xi}_{\mathbf{k}}^{(r)}(t)  &  =u_{\mathbf{k}}^{(r)}(\{\hat{\xi
}\},t)dt+\sqrt{\frac{2\nu}{\Delta^{3}\mathbf{k}}}dW_{\mathbf{k}}%
^{(r)}(t)~~~~(dt>0),\label{dgzi+}\\
d\hat{\xi}_{\mathbf{k}}^{(r)}(t)  &  =\tilde{u}_{\mathbf{k}}^{(r)}(\{\hat{\xi
}\},t)dt+\sqrt{\frac{2\nu}{\Delta^{3}\mathbf{k}}}d\tilde{W}_{\mathbf{k}}%
^{(r)}(t)~~~~(dt<0). \label{dgzi-}%
\end{align}
\end{subequations}
The unknown functionals $u_{\mathbf{k}}^{(r)}(\left\{  {\xi}\right\}  ,t)$ and
$\tilde{u}_{\mathbf{k}}^{(r)}(\left\{  {\xi}\right\}  ,t)$ of $\left\{  {\xi
}\right\}  =\left\{  {\xi}_{\mathbf{k}}^{(r)},r=1,2,\ \forall\mathbf{k}%
\right\}  $ are determined by the variational procedure. The noise terms
$dW_{\mathbf{k}}^{(r)}(t)$ and $d\tilde{W}_{\mathbf{k}}^{(r)}(t)$ in the above
are defined by the two sets of independent Wiener processes,
\begin{subequations}
\begin{align}
E[dW_{\mathbf{k}}^{(r)}(t)]  &  =0,\\
E[dW_{\mathbf{k}}^{(r)}(t)dW_{\mathbf{k}^{\prime}}^{(s)}(t)]  &  =\delta
_{rs}\delta_{\mathbf{k,k^{\prime}}}^{(3)}|dt|,
\end{align}
\end{subequations}
Another one $d\tilde{W}_{\mathbf{k}}^{(r)}$ satisfies the same correlation
property and all other correlations vanish.

The functional relation between $u_{\mathbf{k}}^{(r)}$ and $\tilde
{u}_{\mathbf{k}}^{(r)}$ is called the consistency condition and given by
\begin{equation}
u_{\mathbf{k}}^{(r)}(\{{\xi}\},t)=\tilde{u}_{\mathbf{k}}^{(r)}(\{\xi
\},t)+\frac{2\nu}{\Delta^{3}\mathbf{k}}\frac{\partial}{\partial{\Xi
}_{\mathbf{k}}}\ln\rho(\{{\xi}\},t), \label{consis-k}%
\end{equation}
where $\rho$ represents the probability density of the configuration of the
stochastic filed $\hat{\xi}_{\mathbf{k}}$, and defined by
\begin{equation}
\rho(\{\xi\},t)=E\left[  \prod_{r=1}^{2}\prod_{\mathbf{k}}\delta({\xi
}_{\mathbf{k}}^{(r)}-\hat{\xi}_{\mathbf{k}}^{(r)}(t))\right]  .
\end{equation}
The dynamics of $\rho$ is obtained by the functional Fokker-Planck equation
as
\begin{equation}
\partial_{t}\rho(\{\xi\},t)=-\sum_{r,\mathbf{k}}\frac{\partial}{\partial
\xi_{\mathbf{k}}^{(r)}}(\rho(\{\xi\},t)U_{\mathbf{k}}^{(r)}(\{\xi\},t)),
\label{fp-k}%
\end{equation}
where
\begin{equation}
U_{\mathbf{k}}^{(r)}(\{\xi\},t)=\frac{u_{\mathbf{k}}^{(r)}(\{\xi
\},t)+\tilde{u}_{\mathbf{k}}^{(r)}(\{\xi\},t)}{2}.
\end{equation}

The stochastic action which we should optimize is expressed as
\begin{equation}
I_{sto}  =-\frac{1}{8}\int_{t_{i}}^{t_{f}}dt(\Delta^{3}\mathbf{x}%
)\sum_{\mathbf{x}}E[\hat{F}_{+}^{\mu\nu}(\hat{F}_{+})_{\mu\nu}+\hat{F}%
_{-}^{\mu\nu}(\hat{F}_{-})_{\mu\nu}], \label{sto_act}%
\end{equation}
with $\Delta^{3}\mathbf{x}=(\Delta x)^{3}$. Here $\hat{F}_{\pm}^{\mu\nu}$ is
obtained from Eq.(\ref{Fmunu+}) and Eq.(\ref{Fmunu-}) by substituting
stochastic variables and replacing
\begin{equation}
(\partial_{t})_{+}\longrightarrow D,~~~~(\partial_{t})_{-}\longrightarrow
\tilde{D},
\end{equation}
where $D$ and $\tilde{D}$ are the mean forward and backward derivatives,
respectively \cite{kk4}.

The stochastic variation for the transverse component $\hat{\xi}_{\mathbf{k}%
}^{(r)}$ leads to
\begin{align}
\left(  \partial_{t}+\sum_{r^{\prime},\mathbf{k}}U_{\mathbf{k^{\prime}}%
}^{(r^{\prime})}(\{\xi\},t)\frac{\partial}{\partial\xi_{\mathbf{k^{\prime}}%
}^{(r^{\prime})}}\right)  U_{\mathbf{k}}^{(r)}(\{\xi\},t)  &  =-\frac{1}%
{2}\frac{\partial}{\partial\xi_{\mathbf{k}}^{(r)}}\sum_{r^{\prime
},\mathbf{k^{\prime}}}\omega_{\mathbf{k^{\prime}}}^{2}(\xi_{\mathbf{k^{\prime
}}}^{(r^{\prime})})^{2}\nonumber\\
&  \hspace{-4cm}+\frac{2\nu^{2}}{(\Delta^{3}\mathbf{k})^{2}}\frac{\partial
}{\partial\xi_{\mathbf{k}}^{(r)}}\left\{  \rho^{-1/2}(\{\xi\},t)\sum
_{r^{\prime},\mathbf{k^{\prime}}}\left(  \frac{\partial}{\partial
\xi_{\mathbf{k^{\prime}}}^{(r^{\prime})}}\right)  ^{2}\rho^{1/2}%
(\{\xi\},t)\right\}  , \label{euler-k}%
\end{align}
where $\omega_{\mathbf{k}}=c|\mathbf{q}_{\mathbf{k}}|$. The variations for the
temporal and longitudinal components lead to other equations, and will be
discussed in Sec. \ref{sec:gauge}.

The set of equations~~(\ref{fp-k}) and~(\ref{euler-k}) can be cast into the
form of the functional Schr\"{o}dinger equation,
\begin{equation}
i\hbar\partial_{t}\Psi(\{\xi\},t)=H\Psi(\{\xi\},t), \label{fse}%
\end{equation}
where
\begin{equation}
H=(\Delta^{3}\mathbf{k})\sum_{r,\mathbf{k}}\left[  -\frac{\hbar^{2}c^{2}%
}{2(\Delta^{3}\mathbf{k})^{2}}\left(  \frac{\partial}{\partial\xi_{\mathbf{k}%
}^{(r)}}\right)  ^{2}+\frac{1}{2c^{2}}\omega_{\mathbf{k}}^{2}(\xi_{\mathbf{k}%
}^{(r)})^{2}\right]  .
\end{equation}
Here we set $\nu=\hbar c^{2}/2$. The wave functional is defined by
\begin{equation}
\Psi(\{\Xi\},t)=\sqrt{\rho(\{\xi\},t)}e^{i\theta(\{\Xi\},t)},
\end{equation}
with the phase introduced by
\begin{equation}
U_{\mathbf{k}}^{(r)}(\{\xi\},t)=\frac{\hbar c^{2}}{\Delta^{3}\mathbf{k}}%
\frac{\partial}{\partial\xi_{\mathbf{k}}^{(r)}}\theta(\{\xi\},t).
\end{equation}

All Fock state vectors are given by the stationary solutions of the functional
Schr\"{o}dinger equation. In particular, the vacuum state is given by
\begin{equation}
\Psi_{vac}(\{\xi\})=\prod_{\mathbf{k}}\sqrt{\frac{\omega_{\mathbf{k}}%
(\Delta^{3}\mathbf{k})}{\hbar c^{2}\pi}}e^{-(\Delta^{3}\mathbf{k}%
)\omega_{\mathbf{k}}((\xi_{\mathbf{k}}^{(1)})^{2}+(\xi_{\mathbf{k}}^{(2)}%
)^{2})/(2\hbar c^{2})},
\end{equation}
which is normalized by one. Equivalently, by using Eq. (\ref{xi}), this can be
expressed as a functional of $A_{\mathbf{k}}^{(r)}(t)$ as
\begin{equation}
\Psi_{vac}(\{R,I\})=\prod_{\mathbf{k}}\frac{\omega_{\mathbf{k}}(\Delta
^{3}\mathbf{k})}{\hbar c^{2}\pi}e^{-(\Delta^{3}\mathbf{k})\omega_{\mathbf{k}%
}\sum_{r=1,2}((R_{\mathbf{k}}^{(r)})^{2}+(I_{\mathbf{k}}^{(r)})^{2})/(4\hbar
c^{2})}.\label{vac2}%
\end{equation}
It is however noted that, for example, $R_{\mathbf{k}}^{(r)}$ and
$R_{\mathbf{-k}}^{(r)}$ are not independent in the expression~(\ref{vac2}).

All other stationary states of the functional Schr\"{o}dinger equation are
obtained by operating creation operators to this vacuum state. The creation
and annihilation operators in this case are defined by
\begin{subequations}
\begin{align}
a_{\mathbf{k},r} + a_{\mathbf{-k},r}  &  = l_{\mathbf{k}} R^{(r)}_{\mathbf{k}}
+ \frac{1}{l_{\mathbf{k}}} \frac{\partial}{\partial R^{(r)}_{\mathbf{k}}} ,\\
a^{\dagger}_{\mathbf{k},r} + a^{\dagger}_{\mathbf{-k},r}  &  = l_{\mathbf{k}}
R^{(r)}_{\mathbf{k}} - \frac{1}{l_{\mathbf{k}}} \frac{\partial}{\partial
R^{(r)}_{\mathbf{k}}} ,\\
a_{\mathbf{k},r} - a_{\mathbf{-k},r}  &  = i \left(  l_{\mathbf{k}}
I^{(r)}_{\mathbf{k}} + \frac{1}{l_{\mathbf{k}}} \frac{\partial}{\partial
I^{(r)}_{\mathbf{k}}} \right)  ,\\
a^{\dagger}_{\mathbf{k},r} - a^{\dagger}_{\mathbf{-k},r}  &  = -i \left(
l_{\mathbf{k}} I^{(r)}_{\mathbf{k}} - \frac{1}{l_{\mathbf{k}}} \frac{\partial
}{\partial I^{(r)}_{\mathbf{k}}} \right)  ,
\end{align}
\end{subequations}
where $l_{\mathbf{k}} = \sqrt{\frac{\omega_{\mathbf{k}}}{\hbar c^{2}}
\Delta^{3} \mathbf{k}}$.

Using these expressions, the Hamiltonian operator can be expressed as
\begin{equation}
H = \sum_{r=1}^{2}\sum_{\mathbf{k}} \frac{\hbar\omega_{\mathbf{k}}}{2}
(a^{\dagger}_{\mathbf{k},r}a_{\mathbf{k},r} + a_{\mathbf{k},r}a^{\dagger
}_{\mathbf{k},r}).
\end{equation}
Moreover, by using the definition given in Ref.~\cite{kk4}, the propagator
described by this functional Schr\"{o}dinger equation is given by
\begin{equation}
\Delta^{ij}_{\bot} (x) = \frac{1}{2\pi V} \int dk^{0} \sum_{\mathbf{k}}
\left(  \delta_{ij} - \frac{(q_{\mathbf{k}})^{i}(q_{\mathbf{k}})^{j}%
}{\mathbf{q}_{\mathbf{k}}^{2}}\right)  \frac{e^{-ikx}}{k^{2} + i\epsilon},
\label{propa}%
\end{equation}
where $kx = k^{0} ct - \mathbf{k}\cdot\mathbf{x}$ and $k^{2} = k^{\mu}k_{\mu}=
(k^{0})^{2} - \mathbf{k}^{2}$. In the continuum limit where $\mathbf{q}%
_{\mathbf{k}} = \mathbf{k}$, these are exactly the same as those in the
canonical quantization with the Coulomb gauge condition, although we still
have equations for the temporal and longitudinal components in our formulation.

For the sake of later discussion, it should be remembered that the above
propagator coincides with Green's function satisfying,
\begin{equation}
\left(  \frac{1}{c^{2}}\partial_{t}^{2}-\Delta_{\mathbf{x}}\right)
\Delta_{\bot}^{ij}(x-x^{\prime})=-\left(  \delta_{ij}-\frac{\nabla_{i}%
\nabla_{j}}{\Delta_{\mathbf{x}}}\right)  \delta^{(4)}(x-x^{\prime}).
\end{equation}

\subsection{temporal and longitudinal components and gauge fixing conditions}

\label{sec:gauge}

The optimized solutions of the temporal and longitudinal components are given
by the variations of these components for the stochastic action~(\ref{sto_act}%
), and we obtain
\begin{subequations}
\begin{align}
\left.  \left[  (\mathbf{e}_{\mathbf{k},3}\cdot i\mathbf{q}_{\mathbf{-k}%
})DA_{\mathbf{k}}^{(3)}(t)+(\mathbf{e}_{\mathbf{k},3}\cdot i\mathbf{q}%
_{\mathbf{-k}})\tilde{D}A_{\mathbf{k}}^{(3)}(t)+2c\mathbf{q}_{\mathbf{k}}%
^{2}A_{\mathbf{k}}^{(0)}(t)\right]  _{\hat{\xi}=\Xi}\right.   &
=0,\label{temp}\\
\left.  \left[  (\tilde{D}D+D\tilde{D})A_{\mathbf{k}}^{(3)}(t)-c(\mathbf{e}%
_{\mathbf{-k},3}\cdot i\mathbf{q}_{\mathbf{k}})\tilde{D}A_{\mathbf{k}}%
^{(0)}(t)-c(\mathbf{e}_{\mathbf{-k},3}\cdot i\mathbf{q}_{\mathbf{k}%
})DA_{\mathbf{k}}^{(0)}(t)\right]  _{\hat{\xi}=\Xi}\right.   &  =0.
\label{long}%
\end{align}
\end{subequations}
In the present calculation, fluctuations are introduced only through the
transverse components, and these two equations do not have any term depending
on $\hat{\xi}_{\mathbf{k}}^{(r)}$. Thus $A_{\mathbf{x}}^{0}(t)$ and
$\mathbf{A}_{\Vert\mathbf{x}}(t)$ are deterministic quantities. Then the mean
forward and backward derivatives are, in the end, reduced to the partial time
derivative and the above equations are simplified as
\begin{subequations}
\begin{align}
(\mathbf{e}_{\mathbf{k},3}\cdot i\mathbf{q}_{\mathbf{-k}})\partial
_{t}A_{\mathbf{k}}^{(3)}(t)+c\mathbf{q}_{\mathbf{k}}^{2}A_{\mathbf{k}}%
^{(0)}(t)  &  =0,\label{temp-1}\\
\partial_{t}^{2}A_{\mathbf{k}}^{(3)}(t)-c(\mathbf{e}_{\mathbf{-k},3}\cdot
i\mathbf{q}_{\mathbf{k}})\partial_{t}A_{\mathbf{k}}^{(0)}(t)  &  =0.
\label{long-1}%
\end{align}
\end{subequations}
These two equations are essentially equivalent, and we need to introduce an
additional condition to determine uniquely $A_{\mathbf{k}}^{(0)}$ and
$A_{\mathbf{k}}^{(3)}$ as is well-known in classical electromagnetism. In the
following, we consider the Coulomb and Lorentz gauge conditions.

\subsubsection{Coulomb gauge}

The Coulomb gauge condition $\nabla\cdot\mathbf{A}_{\mathbf{x}}(t)=0$, leads
immediately to $\mathbf{A}_{\Vert\mathbf{x}}(t)=0$, that is,
\begin{equation}
A_{\mathbf{k}}^{(3)}(t)=0.
\end{equation}
Substituting this result to the equations derived from the variation, we find
that the temporal component also vanishes,
\begin{equation}
A_{\mathbf{k}}^{(0)}(t)=0.
\end{equation}
That is, the temporal and longitudinal components completely disappear and we
only need to consider the transverse components.

In short, the result of the canonical quantization with the Coulomb gauge
condition is completely reproduced in this choice of the gauge condition.

\subsubsection{Lorentz gauge}

As was discussed, we can treat $A_{\mathbf{k}}^{(0)}$ and $A_{\mathbf{k}}^{3}$
are c-number fields and thus the form of the Lorentz gauge condition is
well-known as
\begin{equation}
\partial_{t}A_{\mathbf{x}}^{0}(t)+c\nabla\cdot\mathbf{A}_{\Vert\mathbf{x}%
}(t)=0,
\end{equation}
or equivalently,
\begin{equation}
\partial_{t}A_{\mathbf{k}}^{(0)}(t)-c(\mathbf{e}_{\mathbf{k},3}\cdot
i\mathbf{q}_{\mathbf{-k}})A_{\mathbf{k}}^{(3)}(t)=0.
\end{equation}
Substituting this into Eqs.~(\ref{temp}) and~(\ref{long}), we find that the
temporal and longitudinal components are, respectively, given by the solutions
to the following equations,
\begin{subequations}
\begin{align}
(\partial_{t}^{2}+\omega_{\mathbf{k}}^{2})A_{\mathbf{k}}^{(0)}(t)  &  =0,\\
(\partial_{t}^{2}+\omega_{\mathbf{k}}^{2})A_{\mathbf{k}}^{(3)}(t)  &  =0.
\end{align}
\end{subequations}
Green's functions for these c-number fields are then
\begin{align}
\left(  \frac{1}{c^{2}}\partial_{t}^{2}-\Delta_{\mathbf{x}}\right)
\Delta^{00}(x-x\prime)  &  =\delta^{(4)}(x-x^{\prime}),\\
\left(  \frac{1}{c^{2}}\partial_{t}^{2}-\Delta_{\mathbf{x}}\right)
\Delta_{\Vert}^{ij}(x-x^{\prime})  &  =-\frac{\nabla_{i}\nabla_{j}}%
{\Delta_{\mathbf{x}}}\delta^{(4)}(x-x^{\prime}).
\end{align}
Together with Eq.~(\ref{propa}), Green's functions are expressed in a unified
way as
\begin{equation}
\Delta^{\mu\nu}(x-x^{\prime})=\frac{1}{(2\pi)^{4}}\int d^{4}k\frac{-g^{\mu\nu
}}{k^{2}+i\epsilon}e^{-ik(x-x^{\prime})}.
\end{equation}

This Green function coincides with the covariant expression of the
propagator in the canonical quantization with the Lorentz gauge condition.
However, differently from the case of the Coulomb gauge condition, this result
is not equivalent to that of the usual canonical quantization, since the
temporal and longitudinal components behave as c-number fields and are not
replaced with the stochastic variables. However, the behaviors of the temporal
and longitudinal components are completely changed when there is the coupling
with charged matter fields, as will be shown next.

\subsection{Effect of interaction}

\label{int}

To see how the behaviors of the temporal and longitudinal components are
modified by the effect of interaction, let us consider the coupling with the
complex Klein-Gordon field, which is described by the following stochastic
Lagrangian density,
\begin{align}
\mathcal{L}_{SVM}  &  = -\frac{1}{8} \hat{F}^{\mu\nu}_{+} (\hat{F}_{+}%
)_{\mu\nu} -\frac{1}{8}\hat{F}^{\mu\nu}_{-} (\hat{F}_{-})_{\mu\nu}\nonumber\\
&  + \frac{1}{c^{2}} \frac{(D^{\mu}_{+} \hat{\phi})^{*} ((D_{+})_{\mu}%
\hat{\phi}) + (D^{\mu}_{-} \hat{\phi})^{*}((D_{-})_{\mu}\hat{\phi}) }{2} -
\mu^{2} \hat{\phi}^{*} \hat{\phi},
\end{align}
where
\begin{subequations}
\begin{align}
D_{+}^{\mu}  &  = (D/c - ie A^{0}/(\hbar c), -\nabla- ie \mathbf{A}/(\hbar c)
),\\
D_{-}^{\mu}  &  = (\tilde{D}/c - ie A^{0}/(\hbar c), -\nabla- ie
\mathbf{A}/(\hbar c) ).
\end{align}
\end{subequations}
Then, the stochastic variations of the temporal and longitudinal components
respectively lead to
\begin{align}
\left[  \frac{1}{2}(\nabla D + \nabla\tilde{D}) \cdot\mathbf{A}_{\Vert
\mathbf{x}}(t) + c \Delta_{\mathbf{x}} A^{0}_{\mathbf{x}}(t) + \rho
(\{\hat{\phi}, A^{0} \}) \right]  _{\hat{\phi}_{R} = \phi_{R}, \hat{\phi}_{I}
= \phi_{I}}  & = 0 ,\label{tempo-int}\\
\left[  \frac{1}{2} (\tilde{D}D + D\tilde{D}) \mathbf{A}_{\Vert\mathbf{x}}(t)
+ \frac{c}{2} (\tilde{D}\nabla+ D \nabla)A^{0}_{\mathbf{x}}(t) -
\mathbf{J}(\{\hat{\phi}, \mathbf{A}_{\Vert} \}) \right]  _{\hat{\phi}_{R} =
\phi_{R}, \hat{\phi}_{I} = \phi_{I}}  & = 0. \label{long-int}%
\end{align}
where
\begin{subequations}
\begin{align}
\rho(\{\phi, A^{0}\})  & = \frac{e}{\hbar c^{3}} \left\{  \phi_{R\mathbf{x}%
}(t) \frac{D + \tilde{D}}{2} \phi_{I\mathbf{x}}(t) - \phi_{I\mathbf{x}}(t)
\frac{D + \tilde{D}}{2} \phi_{R\mathbf{x}}(t) - \frac{e}{\hbar}A^{0}%
_{\mathbf{x}}(t) \sum_{i=R,I} \phi_{i\mathbf{x}}^{2} (t) \right\} ,\nonumber\\
\\
\mathbf{J}(\{\phi, \mathbf{A}_{\Vert}\})  & = -\frac{e}{\hbar c} \left\{
\phi_{R\mathbf{x}}(t) \nabla\phi_{I\mathbf{x}}(t) - \phi_{I\mathbf{x}}(t)
\nabla\phi_{R\mathbf{x}} (t) + \frac{e}{\hbar c} \mathbf{A}_{\Vert\mathbf{x}%
}(t) \sum_{i=R,I} \phi_{i\mathbf{x}}^{2} (t) \right\} .
\end{align}
\end{subequations}
Here real stochastic variables are introduced by $\hat{\phi}_{\mathbf{x}} =
(\hat{\phi}_{R\mathbf{x}} + i \hat{\phi}_{I\mathbf{x}})/\sqrt{2}$, following
Ref.~\cite{kk4}. We can show that these $\rho$ and $\mathbf{J}$ satisfies the
equation of continuity by using the stochastic Noether's theorem.

Because now these equations depends on the stochastic variables $\hat{\phi
}_{i\mathbf{x}}(t)$, the two components $A_{\mathbf{x}}^{0}(t)$ and
$\mathbf{A}_{\Vert\mathbf{x}}(t)$ become functionals of the configuration of
the complex Klein-Gordon field $\phi_{i\mathbf{x}}$. As a consequence, $D$ and
$\tilde{D}$ cannot be replaced simply by the partial time derivative, but we
should use the following expressions,
\begin{subequations}
\begin{align}
D  &  =\left(  \partial_{t}+\sum_{i=R,I}\sum_{\mathbf{x}}u_{i,\mathbf{x}%
}^{\phi}(\{\phi\},t)\frac{\partial}{\partial\phi_{i\mathbf{x}}}+\frac{\hbar
c^{2}}{2(\Delta^{3}\mathbf{x})}\sum_{i=R,I}\sum_{\mathbf{x}}\frac{\partial
^{2}}{\partial\phi_{i\mathbf{x}}^{2}}\right)  ,\\
\tilde{D}  &  =\left(  \partial_{t}+\sum_{i=R,I}\sum_{\mathbf{x}}\tilde
{u}_{i,\mathbf{x}}^{\phi}(\{\phi\},t)\frac{\partial}{\partial\phi
_{i\mathbf{x}}}-\frac{\hbar c^{2}}{2(\Delta^{3}\mathbf{x})}\sum_{i=R,I}%
\sum_{\mathbf{x}}\frac{\partial^{2}}{\partial\phi_{i\mathbf{x}}^{2}}\right)  ,
\end{align}
\end{subequations}
where $u_{i,\mathbf{x}}^{\phi}$ and $\tilde{u}_{i,\mathbf{x}}^{\phi}$ are,
respectively, the mean forward and backward derivatives for the charged matter
field $\hat{\phi}_{i\mathbf{x}}$,
\begin{subequations}
\begin{align}
D\hat{\phi}_{i\mathbf{x}}  &  =u_{i,\mathbf{x}}^{\phi}(\{\hat{\phi}\},t),\\
\tilde{D}\hat{\phi}_{i\mathbf{x}}  &  =\tilde{u}_{i,\mathbf{x}}^{\phi}%
(\{\hat{\phi}\},t).
\end{align}
\end{subequations}

In short, the temporal and longitudinal components are generally given by
functionals of the configuration of charged matter fields whose forms are
determined by the stochastic variation. When there is no charged matter, these
components behave as classical fields because of the lack of the source of
fluctuation, and coincide with the classical Maxwell's equations as we have discussed.

\section{Stochastic Variation for Lagrangian with Fermi Term}

\label{sec:fix}

In this section, we apply the SVM quantization to the gauge Lagrangian with
the Fermi term. This formulation in the canonical quantization is known as the
method of Gupta-Bleuler, where the concept of the Fock space is extended by
introducing the indefinite metric. In this section, we reproduce this result
by extending the concept of a stochastic process.

The Lagrangian density is
\begin{equation}
\mathcal{L} = -\frac{1}{8} [ F^{\mu\nu}_{+} (F_{+})_{\mu\nu} + F^{\mu\nu}_{-}
(F_{-})_{\mu\nu}] - \frac{1}{4} [(\partial^{\mu}_{+} A_{\mu\mathbf{x}})^{2} +
(\partial^{\mu}_{-} A_{\mu\mathbf{x}})^{2} ] . \label{lag-fermiterm}%
\end{equation}
The second term on the right hand side is the Fermi term. The variable in the
$\mathbf{k}-$representation is given by $A^{(r)}_{\mathbf{k}}$ which is
defined by
\begin{align}
A^{\mu}= \sqrt{\Delta^{3} \mathbf{k}} \sum_{r=0}^{3} \sum_{\mathbf{k}}
\varepsilon_{r\mathbf{k}}^{\mu}A^{(r)}_{\mathbf{k}} (t) \frac{e^{i\mathbf{kx}%
}}{\sqrt{V}}.
\end{align}
Here the covariant polarization vector $\varepsilon_{r\mathbf{k}}^{\mu}$ is
defined by
\begin{equation}
\varepsilon_{0\mathbf{k}}^{\mu} = (1,0,0,0),~~~~ \varepsilon_{r\mathbf{k}%
}^{\mu} = (0, \mathbf{e}_{\mathbf{k},r}),
\end{equation}
and satisfies
\begin{subequations}
\begin{align}
\varepsilon_{r\mathbf{k}}^{\mu}(\varepsilon_{s\mathbf{k}})_{\mu}  &  = -
\zeta_{r} \delta_{rs},\\
\sum_{r=0}^{3} \zeta_{r} (\varepsilon_{r\mathbf{k}}^{\mu})\varepsilon
_{r\mathbf{k}}^{\nu}  &  = - g^{\mu\nu},
\end{align}
\end{subequations}
with $\zeta_{0} = -1$, $\zeta_{1} = \zeta_{2} = \zeta_{3} = 1$.

Substituting these definitions into Eq.~(\ref{lag-fermiterm}), the
corresponding stochastic Lagrangian is given by
\begin{align}
L_{SVM}  &  = -\frac{(\Delta^{3}\mathbf{k})}{4c^{2}} \sum_{r,\mathbf{k}}
(-\zeta_{r}) E[ (D\hat{\xi}^{(r)}_{\mathbf{k}} (t))^{2} + (\tilde{D}\hat{\xi
}^{(r)}_{\mathbf{k}} (t))^{2} - 2\omega_{\mathbf{k}}^{2} (\hat{\xi}%
^{(r)}_{\mathbf{k}} (t))^{2} ],
\end{align}
where we have introduced the real variables as Eq. (\ref{eqn:rea-ima}) for all
$r$'s, and set
\begin{equation}
\xi^{(r)}_{\mathbf{k}} = R^{(r)}_{\mathbf{k}},~~~~\xi^{(r)}_{\mathbf{-k}} =
I^{(r)}_{\mathbf{k}},~~~~(r=0,1,2,3) .
\end{equation}
To obtain this expression, we have dropped the term which is expressed in the
form of the total time derivative by using the stochastic partial integration
formula,
\begin{equation}
E[(D\hat{X}) \hat{Y} + \hat{X} (\tilde{D}\hat{Y})] = \frac{d}{dt}E[\hat{X}%
\hat{Y}].
\end{equation}

Differently from the previous case, the dynamics of this Lagrangian density is
described by the four independent fields and hence we need to introduce four
forward and backward SDEs,
\begin{subequations}
\begin{align}
d\hat{\xi}^{(r)}_{\mathbf{k}}(t) = u^{(r)}_{\mathbf{k}} dt + \sqrt{\frac{\hbar
c^{2}}{\Delta^{3} \mathbf{k}}} dW^{(r)}_{\mathbf{k}}(t)~~~~(dt>0),\\
d\hat{\xi}^{(r)}_{\mathbf{k}}(t) = \tilde{u}^{(r)}_{\mathbf{k}} dt +
\sqrt{\frac{\hbar c^{2}}{\Delta^{3} \mathbf{k}}} d\tilde{W}^{(r)}_{\mathbf{k}%
}(t)~~~~(dt<0),
\end{align}
\end{subequations}
where
\begin{subequations}
\begin{align}
E[dW^{(r)}_{\mathbf{k}}(t)]  &  = 0,\\
E[dW^{(r)}_{\mathbf{k}}(t) dW^{(s)}_{\mathbf{k^{\prime}}}(t)]  &  = \zeta_{r}
\delta_{rs}\delta_{\mathbf{k,k^{\prime}}} |dt|. \label{w02}%
\end{align}
\end{subequations}
The correlation properties for $d\tilde{W}^{(r)}_{\mathbf{k}}$ is the same as
above, but there is no correlation with $dW^{(r)}_{\mathbf{k}}$ as before.

It should be emphasized that $W^{(0)}_{\mathbf{k}}$ cannot be interpreted as
the usual Wiener process, because $\zeta_{0} < 0$. As far as the authors are 
aware, such a stochastic process is not mathematically defined, and the
introduction of such a process seems to be contradict with our naive
intuition. As we will show below, however, this extraordinary stochastic
process corresponds to the indefinite metric in the canonical quantization.
Thus, in the following, we do not discuss this mathematical consistency, but
simply assume that the usual results for the Wiener process such as the Ito
formula, are still applicable even for $W^{(0)}$. Or, we first assume that
$\zeta_{0}$ is a positive value, and substitute $\zeta_{0} = -1$ in the last
step of the calculation. Certainly, we need to understand this issue further.
However, it is beyond the scope of the present exploratory study, and will be
studied in the future.

Then, the corresponding Fokker-Planck equation for the density functional for
the variable $\xi=\left\{  \xi_{\mathbf{k}}^{(r)}\right\}  $ defined before
and the consistency condition are, respectively calculated as
\begin{align}
\partial_{t}\rho &  =\sum_{r,\mathbf{k}}\frac{\partial}{\partial
\xi_{\mathbf{k}}^{(r)}}\left[  -u_{\mathbf{k}}^{(r)}+\zeta_{r}\frac{\nu
}{\Delta^{3}\mathbf{k}}\frac{\partial}{\partial\xi_{\mathbf{k}}^{(r)}}\right]
\rho=\sum_{r,\mathbf{k}}\frac{\partial}{\partial\xi_{\mathbf{k}}^{(r)}}\left[
-\tilde{u}_{\mathbf{k}}^{(r)}-\zeta_{r}\frac{\nu}{\Delta^{3}\mathbf{k}}%
\frac{\partial}{\partial\xi_{\mathbf{k}}^{(r)}}\right]  \rho,\\
u_{\mathbf{k}}^{(r)}  &  =\tilde{u}_{\mathbf{k}}^{(r)}+\zeta_{r}\frac{2\nu
}{\Delta^{3}\mathbf{k}}\frac{\partial}{\partial\xi_{\mathbf{k}}^{(r)}}\ln\rho,
\end{align}
where
\begin{equation}
\rho(\{\xi\},t)=E\left[  \prod_{r=0}^{3}\prod_{\mathbf{k}}\delta
(\xi_{\mathbf{k}}^{(r)}-\hat{\xi}_{\mathbf{k}}^{(r)})\right]  .
\end{equation}
Note that, in the above results, we obtain an additional $\zeta_{r}$ factor
compared to the previous calculations.

Applying the stochastic variation, we finally obtain the following functional
Sch\"{o}dinger equation,
\begin{equation}
i\hbar\partial_{t}\Psi=(\Delta^{3}\mathbf{k})\sum_{r,\mathbf{k}}\zeta
_{r}\left[  -\frac{\hbar^{2}c^{2}}{2(\Delta^{3}\mathbf{k})^{2}}\left(
\frac{\partial}{\partial\xi_{\mathbf{k}}^{(r)}}\right)  ^{2}+\frac{1}{2c^{2}%
}\omega_{\mathbf{k}}^{2}(\xi_{\mathbf{k}}^{(r)})^{2}\right]  \Psi\equiv H\Psi,
\end{equation}
with $\nu=\hbar c^{2}/2$. Here the wave functional is defined by
$\Psi=\sqrt{\rho}e^{i\theta}$,
with the phase introduced by
\begin{equation}
\frac{u_{\mathbf{k}}^{(r)}(\{\xi\},t)+\tilde{u}_{\mathbf{k}}^{(r)}(\{\xi
\},t)}{2}=\zeta_{r}\frac{\hbar c^{2}}{\Delta^{3}\mathbf{k}}\frac{\partial
}{\partial\xi_{\mathbf{k}}^{(r)}}\theta(\{\xi\},t).
\end{equation}

For $r=1,2$ and $3$, this equation is essentially equivalent to the previous
result~(\ref{fse}), and hence it is already confirmed that a Fock space
equivalent to the canonical quantization can be constructed. Thus, in the
following, we only discuss the temporal component $r=0$.

The Fock state vector for $r=0$ is still defined by the stationary solution of
the functional Schr\"{o}dinger equation. The vacuum state for $r=0$ is then
given by
\begin{equation}
\Psi_{vac}^{r=0}=N^{(0)}\prod_{\mathbf{k}}\exp{\ \left(  \frac{\omega
_{\mathbf{k}}\Delta^{3}\mathbf{k}}{4\hbar c^{2}}\{(R_{\mathbf{k}}^{(0)}%
)^{2}+(I_{\mathbf{k}}^{(0)})^{2}\}\right)  },
\end{equation}
where $N^{(0)}$ is the normalization factor and $R_{\mathbf{k}}^{(0)}$ and
$I_{\mathbf{k}}^{(0)}$ represent the real and imaginary part of $A_{\mathbf{k}%
}^{(0)}$, respectively as is defined in Eq. (\ref{eqn:rea-ima}). Note that the
state vectors constructed here is not normalizable, and to determine the
normalization factor $N^{(0)}$, a certain cutoff should be introduced.

The corresponding creation and annihilation operators are defined by
\begin{subequations}
\begin{align}
a_{\mathbf{k},0}+a_{\mathbf{-k},0}  &  =l_{\mathbf{k}}R_{\mathbf{k}}%
^{(0)}-\frac{1}{l_{\mathbf{k}}}\frac{\partial}{\partial R_{\mathbf{k}}^{(0)}%
},\\
a_{\mathbf{k},0}^{\dagger}+a_{\mathbf{-k},0}^{\dagger}  &  =l_{\mathbf{k}%
}R_{\mathbf{k}}^{(0)}+\frac{1}{l_{\mathbf{k}}}\frac{\partial}{\partial
R_{\mathbf{k}}^{(0)}},\\
a_{\mathbf{k},0}-a_{\mathbf{-k},0}  &  =-i\left(  l_{\mathbf{k}}I_{\mathbf{k}%
}^{(0)}-\frac{1}{l_{\mathbf{k}}}\frac{\partial}{\partial I_{\mathbf{k}}^{(0)}%
}\right)  ,\\
a_{\mathbf{k},0}^{\dagger}-a_{\mathbf{-k},0}^{\dagger}  &  =i\left(
l_{\mathbf{k}}I_{\mathbf{k}}^{(0)}+\frac{1}{l_{\mathbf{k}}}\frac{\partial
}{\partial I_{\mathbf{k}}^{(0)}}\right)  .
\end{align}
\end{subequations}
Then one can easily see that $a_{\mathbf{k}}\Psi_{vac}^{r=0}=0$. Thus other
stationary states associated with $r=0$ are given by applying these operators
to the vacuum state $\Psi_{vac}^{r=0}$.

As a result, the correlations of creation-annihilation operators introduced
for this calculation are summarized as
\begin{equation}
\lbrack a_{\mathbf{k},r},a_{\mathbf{k}^{\prime},s}^{\dagger}]=\zeta_{r}%
\delta_{r,s}\delta_{\mathbf{k,k^{\prime}}},
\end{equation}
and then the Hamiltonian operator is re-expressed as
\begin{equation}
H=\sum_{r=0}^{3}\sum_{\mathbf{k}}\zeta_{r}\frac{\hbar\omega_{\mathbf{k}}}%
{2}(a_{\mathbf{k},r}^{\dagger}a_{\mathbf{k},r}+a_{\mathbf{k},r}a_{\mathbf{k}%
,r}^{\dagger}).
\end{equation}
These are the well-known expressions in the canonical quantization. One can
see that the norm of the state vector related to $a_{\mathbf{k},0}$ can be
negative and this energy is not bounded from below, because of $\zeta_{0}=-1$.
In the canonical quantization, these behaviors are interpreted as the effect
of the indefinite metric. In other word, the effect of the indefinite metric
can be reproduced by introducing a singular stochastic variable which has a
negative second order correlation for the quantization of the temporal
component in SVM.

As is well-known in the canonical quantization, this unphysical behavior is
projected out from the physical state vector by requiring the condition,
$(k^{0}a_{\mathbf{k},0}^{\dagger}+|\mathbf{q}_{\mathbf{k}}|a_{\mathbf{k}%
,3}^{\dagger})\Psi_{phys}=0$.

\section{Concluding Remarks}

\label{sec:conc}

In this paper, we have investigated two aspects inherent to the gauge field
quantization within the framework of SVM quantization scheme.

We first investigated the applicability of the SVM field quantization to the
gauge invariant Lagrangian of electromagnetic fields. We verified that the
quantized dynamics obtained by the stochastic variation still has symmetry
associated with the gauge transform. When the Coulomb gauge condition is
employed, the result of the canonical quantization is reproduced. On the other
hand, when the Lorentz gauge condition is applied, the temporal and
longitudinal components of the gauge field can fluctuate only as the influence
of quantized charged matter fields coupled to electromagnetic fields. This is
different from the well-known result of the canonical quantization, that is,
the Gupta-Bleuler formulation where the temporal and longitudinal components
fluctuate even if there is no interaction. Nevertheless, the well-known
propagator of the canonical quantization is still reproduced in the form of 
Green's functions.

The path integral approach is considered as another
quantization scheme based on the Lagrangian. 
In this approach, it is necessary to introduce 
a certain gauge fixing term to avoid
infinitely many equivalent trajectories associated with the gauge symmetry. 
In addition, the stochastic quantization by
Parisi-Wu was originally proposed as a quantization method without fixing the gauge condition, 
but, as pointed out in Ref. \cite{namiki}, 
one of the gauge conditions is implicitly fixed in their formulation. 
Thus, to embrace our quantization in SVM,
it is worth to study carefully consequences of the gauge conditions in
the SVM quantization scheme from various points of view.

For this purpose, for example, the effect of the interaction with charged
matter fields should be investigated in detail, although it was partly discussed in Sec. \ref{int}. 
Then, we will choose one of the gauge conditions as is done in Sec. \ref{sec:gauge}. 
It is, however, not clear how
the gauge transform is defined for quantities given by the functionals of the
charged matter fields. This may correspond to the introduction of the BRST
transform to the framework of SVM, which is considered to be the gauge
transform for quantized systems.

At the same time, it is interesting to examine whether our result can
be reproduced in the canonical quantization. 
Then the temporal and longitudinal components will be given by the complex 
functions of the operators of the transverse components. 
In this case, to avoid the ambiguity for the order of operators, a certain procedure such as the normal
ordering product will be introduced, meanwhile, our approach do not have 
the ordering of variables because stochastic variables are commutable.

As for Green's functions, Feynman's causal boundary condition was used to
derive $\Delta^{00}$ and $\Delta_{\Vert}^{ij}$, but the use of such a boundary
condition is unusual in the classical electrodynamics. 
As far as the authors are aware, the same causal boundary condition is used in the
absorber theory of Wheeler and Feynman, where the
classical electrodynamics is re-formulated, looking for the
classical origin of the infinite self-energy in Quantum
Electrodynamics \cite{quantumvacuum}.

We further investigated that the concept of a stochastic process is
extended to reproduce the indefinite metric for the purpose of reproducing the
Gupta-Bleuler formulation in SVM. This stochastic process
is similar to the Wiener process, but with a negative second-order
correlation. By introducing this stochastic process, we could reproduce
negative norm states and unbounded energy spectra induced by the indefinite
metric. We need to understand more the mathematical meaning and possibility
for such an extension. This is left as a future task.

In this paper, we have discussed the Coulomb and Lorentz gauge conditions, but
there are still different choices of the gauge conditions, for example, the
Landau gauge condition. To deal with these conditions more systematically, it
may be promising to use an auxiliary field as in the theory proposed by
Nakanishi and Lautrap~\cite{nakanishi}. To investigate this aspect in SVM, it
is necessary to develop the formulation of fermionic degrees of freedom, which
is still left as an open question.

\vspace{1cm}

T. Koide thanks for useful comments of C. E. Aguiar about Ref.
\cite{quantumvacuum}. T. Koide and T. Kodama acknowledge the finantial
supports from CNPq, PRONEX, and FAPERJ. \ K. Tsushima is supported by the
Brazilian Ministry of Science, Technology and Innovation (MCTI-Brazil), and
Conselho Nacional de Desenvolvimento Cient{\'{\i}}fico e Tecnol\'{o}gico
(CNPq), project 550026/2011-8.

\end{document}